\documentclass{SCTYE}

\usepackage{multicol}

\def\vec#1{\ensuremath{\mathbf{#1}}}

 \newcommand{\funits}{~$\rm erg~cm^{-2}~s^{-1}$}
 \newcommand{\lengthunits}{~$R_\odot$}
 \newcommand{\velunits}{~$\rm km~s^{-1}$}
 \newcommand{\nofluxunits}{~$\rm cm^{2}~s^{-1}$}

\newcommand{\aap}{Astron Astrophys}

\newcommand{\jgr}{J Geophys Res}
\newcommand{\apj}{Astrophys J}

\newcommand{\apjs}{Astrophys J Suppl Series}

\newcommand{\solphys}{Sol Phys}

\newcommand{\ssr}{Space Sci Rev}
\newcommand{\grl}{Geophys Res Lett}

\begin{document}

\begin{picture}(0,0){\rm
\put(0,4){\makebox[160truemm][l]{\sf  ?}}}
\end{picture}

\begin{picture}(0,0){\rm
\put(0,36){\makebox[178truemm][l]{\textcolor[rgb]{0.39,0.39,0.39}{\xiaosihao
\sf Article\hfill
\includegraphics{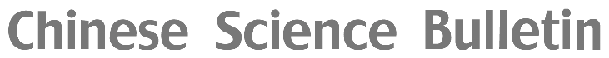}}}}}
\end{picture}

\def\bm{\boldsymbol}

\def\dl{\displaystyle}
\def\wyl{\end{multicols}\end{document}}
\newcommand{\pratio}{\sigma}

\Year{?} %
\Month{?}
\Vol{?} %
\No{?} %
\BeginPage{1} %
\EndPage{?} %
\AuthorMark{{{\rm Li B,} et al.}}
\DOI{?} 

\title{What geometrical factors determine the in situ solar wind speed?}

\author[1,2]{LI Bo}{Corresponding author (email: bbl@sdu.edu.cn)}
\author[1]{CHEN Yao}{}
\author[1]{XIA Li-Dong}{}

\address[{\rm1}]{Shandong Provincial Key Laboratory of Optical Astronomy \& Solar-Terrestrial Environment, \\
 School of Space Science and Physics,
Shandong University at Weihai, 264209 Weihai, PR China;}
\address[{\rm2}]{State Key Laboratory of Space Weather, Chinese Academy of Sciences, 100190 Beijing, PR China}

\maketitle
\vspace{-2mm}{\footnotesize Received ? ; accepted ?}
\vspace*{3mm}


\rule{16.8cm}{0.4pt}\vspace{1mm}\\
\parbox{16.8cm}
{\begin{abstract}
{At present it remains to address why the fast solar wind is fast and the slow wind is slow. Recently we have shown that the field line curvature may substantially influence the wind speed $v$, thereby offering an explanation for the Arge et al. finding that $v$ depends on more than just the flow tube expansion factor. Here we show by extensive numerical examples that the correlation between $v$ and field line curvature is valid for rather general base boundary conditions and for rather general heating functions. Furthermore, the effect of field line curvature is even more pronounced when the proton-alpha particle speed difference is examined. We suggest that any solar wind model has to take into account the field line shape for any quantitative analysis to be made.}
\end{abstract}}

\vspace*{0.35cm} \noindent
\parbox{16.8cm}{\keywords{solar wind, magnetic field, Magnetohydrodynamic waves, space weather}}
\vspace{3mm}

\rule{16.8cm}{0.4pt}\vspace{-0.8mm}\\
\renewcommand{\baselinestretch}{1.2}
\renewcommand{\arraystretch}{1.5}
{\begin{tabular}{lp{0.88\textwidth}}  \scriptsize
{\bf\hspace{-2.6mm} Citation:}\!\!\!\!&\scriptsize Li B et al.
What geometrical factors determine the in situ solar wind speed? Chinese Sci Bull, 201?, ?: 1--?, doi:
?
\\
\end{tabular}}

\rule{16.8cm}{0.4pt}


\textwidth=178truemm \textheight=236truemm

\wuhao\vspace*{5mm}
\begin{multicols}{2}

\renewcommand{\baselinestretch}{1.08} \baselineskip 12.2pt\parindent=10.8pt

\renewcommand{\thefootnote}

\noindent
Half a century after its prediction~\cite{1958ApJ...128..664P} and
   direct detection~\cite{1962Sci...138.1095N},
   how the solar wind is accelerated to its highly supersonic terminal speed
   remains elusive.
Among the extensive experimental findings that provide stringent constraints on these mechanisms,
   those concerning the in situ wind speed $v$ are of primary concern from both
   theoretical and practical standpoints: $v$ and the magnetic field polarity near the Earth
   are the two main parameters involved for predicting geomagnetic activities, a major
   player in space weather~\cite{Zhao_etal_11,WeiY_etal_11, ZongQG_etal_11,FengYY_etal_11}.
A widely used technique to predict $v$ is the Wang-Sheeley-Arge model~\cite{2004JASTP..66.1295A},
   which is based on
   two well-established statistical relations between $v$ and the properties of the flow tube $\cal{L}$
   along which the wind flows.
First, $v$ is inversely correlated with the coronal expansion rate $f_c$
   of $\cal{L}$~\cite{1990ApJ...355..726W}.
Second, for a given $f_c$, $v$ is inversely correlated with the angular distance $\theta_b$ of
   the footpoint of $\cal{L}$ to its nearest coronal hole boundary~\cite{2004JASTP..66.1295A}.
Given the importance of the wind speed prediction, the two results deserve a sound physical explanation.
While the former is well supported
   by a number of wave/turbulence-based solar wind models~(see \cite{2007ApJS..171..520C} and references therein),
   the latter has so far attracted little attention.

Close inspections of coronal images indicate that the
   further away the field line footpoint is from the coronal hole boundary,
   the more curved the field line is (see e.g., the composite image Fig.1 in~\cite{1999SSRv...89....7F}).
Consequently, the effects of $\theta_b$ on the wind speed may be explained in terms of field line curvature.
A detailed modeling study using a sophisticated Alfv\'enic-turbulence-based heating mechanism suggests
   that it is indeed so (\cite{2011A&A...529A.148L}, hereafter paper I).
The model distinguishes between electrons and protons, adopts reasonably complete energy equations including radiative losses,
   electron heat conduction, and solar wind heating.
It also employs the Radiation Equilibrium Boundary (REB) at the base which is set at the mid-Transition region,
   whereby the base electron density automatically adjusts to the downward
   heat flux from the corona~\cite{1988ApJ...325..442W,2001ApJ...546..542L}.
Compared with a straight flow tube with identical radial distribution of the magnetic field strength,
   a curved tube produces a significantly reduced flow speed at 1~AU (by up to $\sim 30\%$),
   accompanied by an enhanced mass flux density.
This effect was interpreted, in view of the general framework by~\cite{1980JGR....85.4681L},
   as a result of the enhanced energy deposition in the near-Sun region 
   in the curved case relative to
   the straight one since the elementary reduction in the wave action flux density
   is a factor of $dl/dr$ larger (Eq.(7) in paper I), where $l$ is the arclength and $r$ the
   heliocentric distance.

The aim of the present paper is to extend paper I by examining whether the effect of field line curvature discovered there
   is robust enough:
   Is it valid only for this particular boundary condition or this particular heating mechanism?
   Will the inclusion of alpha particles, the second most abundant ion species, invalidate the conclusion?
The latter is of particular importance since alpha particles, with their mass being 4 times the proton one,
   may play an important role in regulating the wind parameters~\cite{1992JGR....97.8183L, 1997JGR...10217419L, 2003ApJ...596..621L}.
This paper is organized as follows.
We will give a brief overview of the solar wind model in section~\ref{sec_model_des},
   then present the numerical results in section~\ref{sec_num_res}.
Finally, section~\ref{sec_conc} summarizes the results.

\section{Model Description}
\label{sec_model_des}

Let us start with considering a time-independent solar wind that consists of electrons ($e$), protons ($p$),
  and alpha particles ($\alpha$).
Each species $s$ ($s=e, p, \alpha$) is characterized by its mass $m_s$, electric
  charge $e_s$, number density $n_s$, mass density $\rho_s = n_s m_s$, velocity
  $\vec{v}_s$, temperature $T_s$ and partial pressure $p_s = n_s k_B T_s$.
Here $k_B$ is the Boltzmann constant.
We may express $e_s$ in units of the electron charge $e$, i.e., $e_s = Z_s e$ with $Z_e \equiv -1$
  by definition.
Assuming axisymmetry, quasi-neutrality and quasi-zero current, neglecting solar rotation,
  one may readily decompose the vector equations in the
  standard 5-moment treatment into a force balance equation across the meridional
  magnetic field $\vec{B}$ and a set of transport equations along it
  (for a formal development, see section~2.1 in~\cite{2007ApJ...661.1222L}).
This transport equation set reads
\begin{eqnarray}
&& B\left(\frac{n_k v_k}{B}\right)' =0,
      \label{eq_nk}\\
&& v_{k}v_{k}' +\frac{p_k'}{n_k m_k}
       +\frac{Z_k p_e'}{n_e m_k}+\frac{G M_\odot}{r} (\ln r)'   \nonumber \\
&&     -\frac{1}{n_k m_k}\left(\frac{\delta M_{k}}{\delta t}
             +\frac{Z_k n_k}{n_e}\frac{\delta M_{e}}{\delta t}\right)
       = a_{k}, \label{eq_vk} \\
&& v_s \left(\frac{p_s}{\gamma -1}\right)'
       + \frac{\gamma p_s }{\gamma-1} B\left(\frac{v_{s}}{B}\right)'  \nonumber \\
&&     + B\left(\frac{q_s}{B}\right)'
          -\frac{\delta E_s}{\delta t} -Q_s + L=0,
               \label{eq_ps}
\end{eqnarray}
     where the subscript $s$ refers to all species ($s=e,p,\alpha$),
     while $k$ stands for ion species only ($k = p, \alpha$).
The gravitational constant is denoted by $G$,
     $M_\odot$ is the mass of the Sun.
The prime $'$ denotes the derivative with respect to the arclength $l$,
     measured from the footpoint of a magnetic field line.
The momentum and energy exchange rates due to the Coulomb
     collisions of species $s$ with the remaining ones
     are denoted by $\delta {M}_s/\delta t$ and $\delta E_s/\delta t$,
     respectively.
Their expressions may be found in~\cite{2006A&A...456..359L} for which we use a Coulomb logarithm 
     $\ln\Lambda = 23$.
As for $a_k$, it represents the acceleration exerted on ion species $k$.
Moreover, ${q}_s$ is the heat flux carried by species $s$, and
     $Q_s$ stands for the heating rate applied to
     species $s$ from non-thermal processes.
$L$ is the radiative loss function involved in the electron energy equation only.
It is assumed to be optically thin
     and follows the analytic form given in~\cite{1978ApJ...220..643R}.
To simplify our treatment, we neglect ion heat fluxes ($q_k = 0$), and
   adopt the Spitzer law for the electron one
   $q_e = -\kappa T_e^{5/2} T_e'$ where
   $\kappa=7.8\times 10^{-7}$~erg~K$^{-7/2}$~cm$^{-1}$~s$^{-1}$~\cite{Spitzer62}.

As detailed in paper I, rather than solving for a fully self-consistent multi-dimensional
     solution, we solve Eqs.(\ref{eq_nk}) to~(\ref{eq_ps})
     on a prescribed magnetic field instead, the reason being that
     in a multi-dimensional model the effects of magnetic field expansion
     and field line curvature cannot be distinguished.
In practice, we employ the analytic model given by~\cite{1998A&A...337..940B} for which
     the reference values $[Q,K,a_1] = [1.5, 1, 1.538]$ are used.
The resulting magnetic field configuration agrees favorably with the coronal
  images obtained with SOHO/LASCO~\cite{1997SoPh..175..667S}, and has been widely seen
  as being representative of
  a minimum solar corona.
In line with the Ulysses measurements, the magnetic field strength at~1~AU is
  essentially latitude-independent and is
  3.5~$\gamma$~\cite{1995GeoRL..22.3317S,2011JASTP..73..277S}.
All the presented results will be computed along the field line that reaches 1~AU
  at $88^\circ$ colatitude, nearly maximizing the effect of field line curvature.

\begin{table*}[tp]
\caption{Combinations of base boundary conditions, electron and proton heating, and inclusion of alpha particles}
\label{tab_choices}
\centering
\begin{tabular}{cccccccccc}
\toprule
{\bf section}
      & \multicolumn{3}{c}{\bf Base Boundary Condition} & \multicolumn{2}{c}{\bf $e^-$ heating}
      & \multicolumn{2}{c}{\bf ion heating} & \multicolumn{2}{c}{\bf $\alpha$-particles included?} \\ \cline{2-10}
      & REB       & TR	   &CB		&Y		& N     	& Wave   	& Empirical   	& Y    		& N \\ \hline
\ref{sec_numres_base}
      & \checkmark & 	   &		&\checkmark	&		&\checkmark	&		&		& \checkmark \\
      & 	       &\checkmark &		&\checkmark	&		&\checkmark	&		&		& \checkmark \\
      & 	       &	   &\checkmark	&		&\checkmark	&\checkmark	&		&		& \checkmark \\ \hline
\ref{sec_numres_p_heating}
      & 		&	   &\checkmark	&		&\checkmark	&		&\checkmark	&		&\checkmark  	\\ \hline
\ref{sec_numres_alpha}
      & 		&	   &\checkmark	&		&\checkmark	&		&\checkmark	&\checkmark	&  	\\
\bottomrule
\end{tabular}

\flushleft{\hskip 2.5cm \footnotesize See text for details.} 
\end{table*}

Specifying the boundary condition (BC) at the base $r=1~R_\odot$ is crucial for constructing wind solutions.
We will test three BCs: REB, TR, and CB.
By REB, we mean the Radiation Equilibrium Boundary where
    a base temperature $T_{s0}=5\times 10^5$~K is set,
    the electron density $n_{e0} = 4.8\times 10^3 q_{e0}$, in which the subscript $0$
    represents the base value~\cite{1988ApJ...325..442W, 2001ApJ...546..542L}.
Besides, TR (CB) refers to the case where we set the base at the lower Transition-Region (coronal base),
    $T_{s0} = 10^5$~K ($10^6$~K), the corresponding $n_{e0}$ being $10^9$ ($1.5\times 10^8$)~cm$^{-3}$.

The source terms $a_k$ and $Q_s$ remain to be specified.
In our study they depend on whether alpha particles are included,
    and on how the base BC is implemented.
Electron heating is applied for the REB and TR boundaries to compensate for the significant radiative losses
    at the corresponding temperatures.
It is in the form $Q_{e} = Q_{e0} \exp\left(-l/l_d\right)$, with $Q_{e0}$ (in $10^{-5}$~erg~cm$^{-3}$~g$^{-1}$) and $l_d$ (in $R_\odot$) being
    $1.7$ and $0.06$ for REB,
    $2.9$ and $0.06$ for TR.
For ion heating, either the wave heating scenario or some empirical heating function is employed.
The former choice is used only in the absence of alpha particles.
In this case a wave transport equation (Eq.(4) in paper I) is included where the parallel propagating Alfven waves
    lose their energy via both doing work on the flow throught the ponderomotive force ($a_k$, see Eq.(2) in paper I)
    and dissipation at the Kolmogorov rate.
{(For more recent developments on the role of Magnetohydrodynamic waves in
    energizing plasmas, see e.g., \cite{ChenWu_11}.)}
We use
    a base wave amplitude of $27$, $18$, and $29$\velunits,
    a base value of correlation length being 1.5, 2, and 2$\times 10^4$~km,
    for the REB, TR and CB boundaries, respectively.
These parameters are within observational limits, and produce solar wind solutions  whose terminal speed and proton flux
    agree with in situ measurements.
When ions are heated in an empirical fashion, we assume the external energy comes from an ad hoc energy flux dissipated at
    a constant length $l_{\mathrm{ad}}$,
    the resulting total heating rate being $Q=F_E B/(B_E l_{\mathrm{ad}}) \exp\left(-l/l_{\mathrm{ad}}\right)$.
Here $F_E$ is the input flux scaled to the Earth orbit $r_E = 1$~AU, and $B_E$ is the meridional magnetic field strength at $r_E$.
$Q$ is apportioned between protons and alpha particles, and some energy that alpha particles receive
    turns out necessary to be in the form of momentum deposition,
\begin{eqnarray*}
   && Q_p + \bar{Q}_\alpha = Q, \frac{\bar{Q}_\alpha}{Q_p} =\chi_E\frac{\rho_\alpha}{\rho_p},  \\ \nonumber
   && Q_\alpha + \rho_\alpha v_\alpha a_\alpha = \bar{Q}_\alpha, \frac{a_\alpha}{Q_\alpha} =\frac{\chi_D}{\rho_\alpha v_0},
\end{eqnarray*}
    where $v_0 = \sqrt{k_B T_p/m_p}$ is some characteristic speed.
The momentum deposition to protons is neglected.
This way for empirically heating the ion fluids seems rather involved, but actually it manages to capture some
    essential features of wave heating in a multifluid treatment (for details, see~\cite{2007ApJ...661..593L}, section 3.1).
In practice, we use $F_e=1.4$\funits\ and $l_{\mathrm{ad}}=1.75$\lengthunits, regardless of the presence of alpha particles,
    such that the resulting proton flux and speed at 1~AU agree with the in situ measurements.
When alpha particles exist, we use $\chi_E=6$ and $\chi_D=2$, the purpose being to generate a wind profile
    that corresponds to a positive proton-alpha speed difference (see section~\ref{sec_numres_alpha} for further details.)

Table~1 summarizes our computations. 
The first column indicates in which section the combinations of BCs and ion/electron heating in the same row are tested. 
For boundary conditions imposed at the base, REB (Radiation Equilibrium Boundary), TR (Transition Region), and CB (Coronal Base)
   conditions are experimented with.
For electron heating, Y (N) refers to the case where the heating is switched on (off).
As for the ion heating, ``Wave'' and ``empirical'' represent the case where the ions are heated by turbulent Alfv\'en waves
   and in an empirical fashion, respectively.
The last column indicates whether or not alpha particles are included in the computations,
   with Y and N meaning that they are and are not, respectively.

\centerline{\includegraphics[width=0.95\columnwidth]{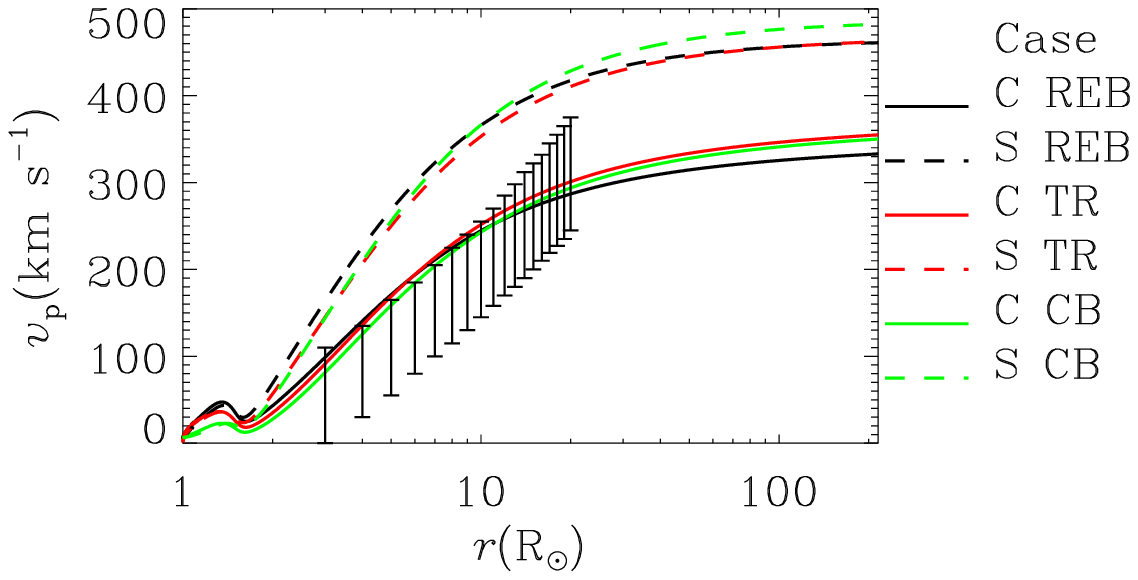}}
\noindent{\footnotesize {\bf Figure 1}\quad Effects of field line curvature on wind speed under
   the influence of different base boundary conditions.
Here the proton speed $v_p$ is shown as a function of heliocentric distance $r$.
The solid (dashed) curves are for the solutions that incorporate (neglect) the field line curvature.
Black, red, and green curves represent solutions using the REB, TR, and CB boundaries, respectively.
Moreover, the vertical bars represent wind speeds measured by tracking small density inhomogeneities
     (blobs) in SOHO/LASCO images as given by~\cite{2000JGR...10525133W}.
}

\section{Numerical Results}
\label{sec_num_res}

Now we are in a position to examine how robust the effects of field line curvature are by varying our choices of
    base BC, the proton heating, and the inclusion of alpha particles.
For any combination of the three choices, we compute two solar wind solutions that differ only in whether or not
    the field line curvature is considered.
If not (yes), the solution is labeled S (C), short for straight (curved).
{Before proceeding, let us briefly discuss what we mean by ``curvature''. 
Similar to paper I, here
  it is defined in an intuitive rather than a precise manner, and refers to how significantly
  the field line differs from a radial shape.
Locally speaking, at any given point along a field line, the curvature can be precisely defined, say, in terms of the
  curvature radius $\cal{R}$.
However, the effect of field line shape on the wind parameters turns out to be an integrated one to which the shape at
  each and every point contributes.
A non-local definition of the field line shape by, say, integrating $1/\cal{R}$ is for sure more precise, but evaluating
  its correlation with the computed wind parameters is  beyond
  the scope of the present paper.
}

\subsection{Influences of the base boundary condition (without alpha particles)}
\label{sec_numres_base}

Let us for now neglect alpha particles but focus on the influence of the boundary condition imposed at the base.
Figure~1 presents the distribution of the proton speed $v_p$ with the heliocentric distance $r$.
It compares the solutions obtained in the curved cases (solid lines) with those in the straight cases (dashed lines).
The black, red, and green curves are for REB, TR, and CB, respectively.
Moreover, the vertical bars represent wind speeds measured by tracking small density inhomogeneities
     (blobs) in SOHO/LASCO images as given by~\cite{2000JGR...10525133W}.
One can see that all lines, both solid and dashed, possess a dip below 2\lengthunits,
    indicating that this local minimum in $v_p$ results from the drastic, lateral expansion of the flow tube rather than
    its curvature.
On the other hand, all the solid curves reproduce the SOHO/LASCO measurements equally well, and they
    reach similar values at 1~AU.
To be specific, the REB, TR, and CB solutions yield a $v_p$ of $333$, $355$, and $351$\velunits, respectively.
The corresponding values in solutions with identical BCs but neglecting field line curvature are
   $461$, $462$, and $482$\velunits\ respectively.
Moreover, the proton flux at 1~AU (in $10^8$\nofluxunits) is found to be $3.65$($3.14$) in the curved (straight) case for REB,
   $3.86$ ($3.47$) for TR,
   $3.57$ ($3$) for CB.
In other words, one finds a reduction in $v_p$, relative to the straight case, of $27.8\%$ for REB,
   $23.2\%$ for TR, and $27.4\%$ for CB,
   and an enhancement in proton flux by $16.1\%$ for REB, $11.4\%$ for TR, and $18.9\%$ for CB, respectively.
Hence, the significant reduction in the terminal proton speed and enhancement in the proton flux
   due to the inclusion of field line curvature
   is not specific to the particular REB condition employed in paper I,
   but happens for rather general boundary conditions.

\subsection{Influences of proton heating}
\label{sec_numres_p_heating}

Is the effect of field line curvature present only for the particular ion heating mechanism used in paper I?
This is examined in Figure~2 where we again neglect alpha particles but do employ a different ion heating function,
   namely the empirical heating rates detailed in section~\ref{sec_model_des}.
Figure~2 presents the radial distributions of the proton speed $v_p$, for both cases C (solid lines)
   and S (dashed lines).
Comparison of Fig.2 with Fig.1 indicates that the proton speed profile in Case C is steeper than in the wave-driven model, nonetheless
   the effect of field line curvature remains the same.
To be specific, the terminal speed is reduced from $583$ in the straight to $399$\velunits in the curved case.
Likewise, the proton flux (in $10^8$\nofluxunits) increases from $2.61$ to $3.47$.
This change in the proton parameters can be readily explained using the same argument as in the wave-driven case.
The energy deposition in the subsonic region can be expressed as an energy flux density scaled to 1~AU, 
   $F_{\mathrm{sub}}=\int_{R_\odot}^{r_C} Q {B_E}/{B} (dl/dl) dr$, evaluating which
   results in
   $F_E [1-\exp(-l_C/l_\mathrm{ad})]$.
Here the subscript $C$ represents the values at the critical point where $v_p$ equals $\sqrt{k_B(T_e + T_p)/m_p}$.
It follows that if in Cases C and S the critical point lies at the same {\it radial} distance,
   $F_{\mathrm{sub}}$ is larger in Case C given that $l_C$ is larger.
And from the general framework presented in~\cite{1980JGR....85.4681L} it can be readily deduced that
   Case C will yield an enhanced proton flux and a reduced proton terminal speed.
Actually this effect is compounded by the fact that in Case C the critical point is located further away
   from the Sun: $r_C = 4.13$ and $3.51$\lengthunits\ in Cases C and S, respectively.

\subsection{Influences of alpha particles}
\label{sec_numres_alpha}

\centerline{\includegraphics[width=0.95\columnwidth]{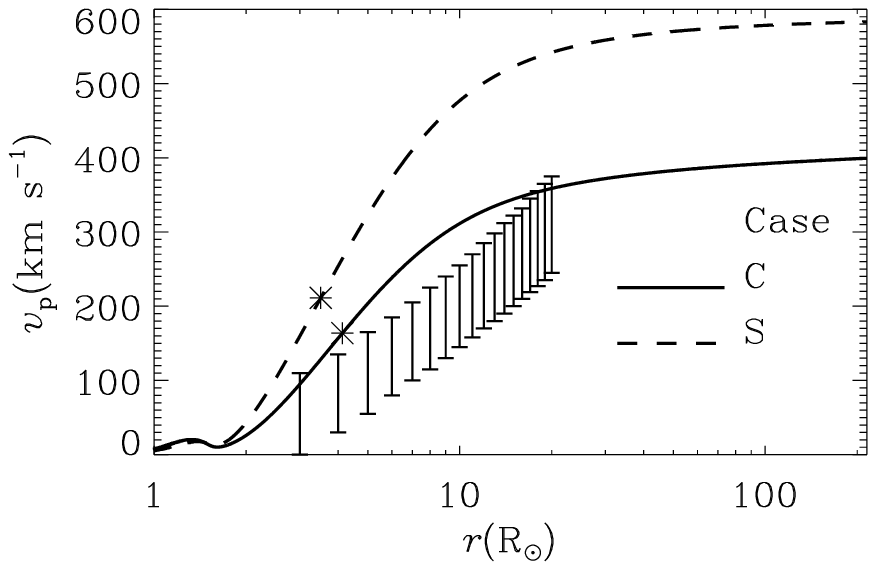}}
\noindent{\footnotesize {\bf Figure 2}\quad Effects of field line curvature on the proton speed for an empirically heated wind.
The solid (dashed) curves are for the solutions that incorporate (neglect) the field line curvature.
The vertical bars have the same meaning as in Figure~1.
As for the asterisks, they represent the critical points where the proton speed equals the sound speed (see text for details).
}

Finally let us address the role of alpha particles.
For simplicity we employ the empirical ion heating rates, since we have already shown that adopting some more sophisticated
    mechanism like wave heating
    will not change our result as far as the effect of field line curvature is concerned.
Figure~3 examines how the flow parameters at 1~AU change with varying $n_{\alpha p, 0}$, the alpha abundance at the coronal base.
These include the proton speed $v_{p, E}$, proton flux density $(n_p v_p)_E$, and proton-alpha
    speed difference $v_{\alpha p, E} = (v_\alpha - v_p)_E$.
The solid and dashed curves are for Cases C and S, respectively.
Figure~3a indicates that for $n_{\alpha p, 0}$ in the range from $3\times 10^{-5}$ to $0.1$,
    the field line curvature consistently reduces the terminal proton speed and enhances the proton flux density.
For $n_{\alpha p, 0} \lesssim 10^{-2}$, introducing alpha particles has little effect on the proton parameters
    and the reduction in $v_{p, E}$ by the field line curvatures is $\sim 183$\velunits, or by $\sim 31.5\%$.
On the other hand, for $n_{\alpha p, 0}=0.1$, Case C yields a $v_{p, E}$ of $331$\velunits,
    or down by $29.6\%$ relative to Case S where $v_{p, E}$ is $470$\velunits.
Hence whether or not the alpha particles are included, introducing field line curvature has
    the same effect on the proton parameters in interplanetary space.

\centerline{\includegraphics[width=0.95\columnwidth]{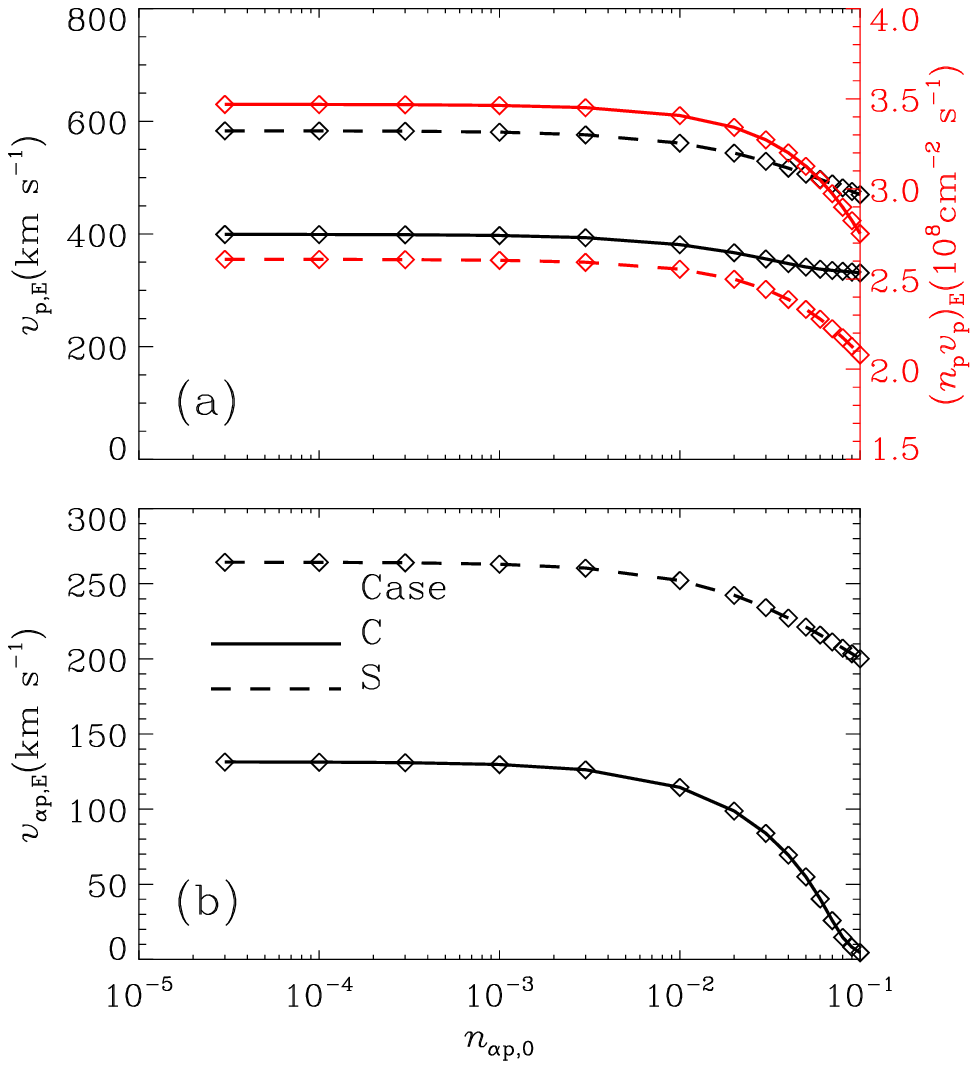}}
\noindent{\footnotesize {\bf Figure 3}\quad
Effects of field line curvature on flow parameters for an electron-proton-alpha wind.
Displayed as a function of the base alpha abundance $n_{\alpha p, 0} = (n_\alpha/n_p)_0$ are values at 1~AU of
   (a) the proton speed $v_{p, E}$ and proton flux density $(n_p v_p)_E$,
  and (b) the proton-alpha speed difference $v_{\alpha p, E} = (v_\alpha - v_p)_{E}$.
The solid (dashed) curves are for the solutions that incorporate (neglect) the field line curvature.
}

The proton-alpha speed difference $v_{\alpha p}$ has been a subject of
    extensive study (e.g., \cite{2003JGRA..108.1036K, 2009PhRvL.102q5001A} and references therein).
Figure~3b shows that the field line curvature significantly reduces $v_{\alpha p}$.
This is true throughout the range of $n_{\alpha p, 0}$ explored.
For instance, the amount of reduction in $v_{\alpha p, E}$
    is $133$\velunits\ for $n_{\alpha p, 0}=3\times 10^{-5}$ (from $264.3$ in Case S to $131.4$\velunits\ in Case C),
    and is $195.5$\velunits\ for $n_{\alpha p, 0}= 0.1$ (from $200$ in Case S to $4.54$\velunits\ in Case C).
This significant reduction in $v_{\alpha p}$ at large distances turns out to be reminiscent of
    what happens in the corona: although not shown, the alpha temperature is substantially higher in Case S than in Case C
    from $1.5$\lengthunits\ onward, resulting in a substantially larger alpha pressure gradient force and
    hence a significantly larger proton-alpha speed difference.    
From this we may conclude that, 
    the effect of field line curvature should be taken into account for a quantitative investigation
    into the low-latitude slow wind,
    and its role in regulating the ion speed difference is even more pronounced. 
Once this effect is introduced, one may expect to see some significant revision to some recent studies (e.g., \cite{2003ApJ...596..621L, 2007A&A...474..997J}), where
    the flow tube is taken to be radially directed.    

\section{Conclusions}
\label{sec_conc}

The present study has been motivated by the fact that for space weather applications, the near-Earth solar wind speed $v$
    is of crucial importance and is being routinely predicted by the Wang-Sheeley-Arge (WSA) model~\cite{2004JASTP..66.1295A}.
This model is based on the relations between $v$ and the coronal expansion rate of the solar wind flow tube~\cite{1990ApJ...355..726W}
    as well as the angular distance of the tube footpoint to its nearest coronal hole boundary~\cite{2004JASTP..66.1295A}. 
While the former correlation has been extensively examined, the latter has been given little physical interpretation, and was
    only recently explained in terms of the field line curvature~\cite{2011A&A...529A.148L}:
    the more curved the flow tube (or equivalently the magnetic field line) is,
    the higher the proton flux density, and the lower the terminal proton speed.
Given that the conclusion was established only via a limited number of computations, and given 
   that providing a sound physical explanation for the WSA model puts considerably more confidence
   in this model,
   we show in this paper by extensive numerical examples that the correlation between
   the terminal proton speed and field line curvature is valid for rather general base boundary conditions
   and for rather general heating functions commonly adopted in the literature. 
Furthermore, its validity is not violated by the introduction of alpha particles, the second most abundant
   ion species in the corona and solar wind.
An immediate consequence of our new computations is that any solar wind model, for the fast and slow wind alike,
   has to take into account the field line shape for any quantitative analysis to be made.
Beside, the effect of field line curvature is even more pronounced when examining the proton-alpha
   particle speed difference, thereby calling for the urgent inclusion of this effect in multi-fluid models
   of the low-latitude solar winds where the ion speed difference is among the primary concerns.

\Acknowledgements{\bahao
This research is supported by the grant NNSFC 40904047,
     and also by the Specialized Research Fund for State Key Laboratories.
YC is supported by grants NNSFC 40825014 and 40890162,
   and LDX by NNSFC 40974105.}


\normalsize \parskip=0mm \baselineskip
18pt\renewcommand{\baselinestretch}{1.1}\footnotesize\parindent=4mm\bahao


\bibliographystyle{unsrt}

\end{multicols}

\vspace{2mm} {\begin{tabular}{lp{0.90\textwidth}}  \scriptsize
{\bf\hspace{-2.6mm} Open Access}&\scriptsize This article is
distributed under the terms of the Creative Commons Attribution
License which permits any use, distribution, and reproduction in any
medium, provided the original author(s) and source are credited.
\end{tabular}}

\end{document}